\RequirePackage{lineno}
\documentclass[aps,prl,onecolumn,amsmath,amssymb,superscriptaddress]{revtex4}
\usepackage{graphicx}
\usepackage{amssymb}
\usepackage{natbib}
\usepackage{color}
\usepackage{lineno}

\newcommand{\beq}{\begin{eqnarray}}
\newcommand{\eeq}{\end{eqnarray}}

\begin{document}
\title{$c$-axis pressure induced antiferromagnetic order in
 optimally P-doped BaFe$_2$(As$_{0.70}$P$_{0.30}$)$_2$ superconductor}

\author{Ding Hu}
\affiliation{Center for Advanced Quantum Studies and Department of Physics, Beijing Normal University, Beijing 100875, China}
\affiliation{Department of Physics and Astronomy, Rice University, Houston, Texas 77005-1827, USA}
\author{Weiyi Wang}
\affiliation{Department of Physics and Astronomy, Rice University, Houston, Texas 77005-1827, USA}
\author{Wenliang Zhang}
\affiliation{Beijing National Laboratory for Condensed Matter Physics, Institute of Physics, Chinese Academy of Sciences, Beijing 100190, China}
\affiliation{University of Chinese Academy of Sciences, Beijing 100049, China}
\author{Yuan Wei}
\affiliation{Beijing National Laboratory for Condensed Matter Physics, Institute of Physics, Chinese Academy of Sciences, Beijing 100190, China}
\affiliation{University of Chinese Academy of Sciences, Beijing 100049, China}
\author{Dongliang Gong}
\affiliation{Beijing National Laboratory for Condensed Matter Physics, Institute of Physics, Chinese Academy of Sciences, Beijing 100190, China}
\affiliation{University of Chinese Academy of Sciences, Beijing 100049, China}
\author{David W. Tam}
\affiliation{Department of Physics and Astronomy, Rice University, Houston, Texas 77005-1827, USA}
\author{Panpan Zhou}
\affiliation{Department of Physics and Astronomy, Rice University, Houston, Texas 77005-1827, USA}
\author{Yu Li}
\affiliation{Department of Physics and Astronomy, Rice University, Houston, Texas 77005-1827, USA}
\author{Guotai Tan}
\affiliation{Center for Advanced Quantum Studies and Department of Physics, Beijing Normal University, Beijing 100875, China}
\author{Yu Song}
\affiliation{Department of Physics and Astronomy, Rice University, Houston, Texas 77005-1827, USA}
\author{Robert Georgii}
\affiliation{Heinz Maier-Leibnitz Zentrum, Technische Universit$\ddot{a}$t M$\ddot{u}$nchen, D-85748 Garching, Germany}
\author{Bj$\rm \ddot{o}$rn Pedersen}
\affiliation{Heinz Maier-Leibnitz Zentrum, Technische Universit$\ddot{a}$t M$\ddot{u}$nchen, D-85748 Garching, Germany}
\author{Huibo Cao}
\affiliation{Neutron Scattering Division, Oak Ridge National Laboratory, Oak Ridge, Tennessee 37831, USA}
\author{Wei Tian}
\affiliation{Neutron Scattering Division, Oak Ridge National Laboratory, Oak Ridge, Tennessee 37831, USA}
\author{Bertrand Roessli}
\affiliation{Laboratory for Neutron Scattering and Imaging, Paul Scherrer Institut, CH-5232 Villigen, Switzerland}
\author{Zhiping Yin}
\affiliation{Center for Advanced Quantum Studies and Department of Physics, Beijing Normal University, Beijing 100875, China}
\author{Pengcheng Dai}
\affiliation{Department of Physics and Astronomy, Rice University, Houston, Texas 77005-1827, USA}
\affiliation{Center for Advanced Quantum Studies and Department of Physics, Beijing Normal University, Beijing 100875, China}

\maketitle
{\bf Superconductivity in BaFe$_2$(As$_{1-x}$P$_{x}$)$_2$ iron pnictides
emerges when its in-plane two-dimensional (2D) orthorhombic lattice distortion associated with nematic phase at $T_s$ and three-dimensional (3D) collinear antiferromagnetic (AF) order at $T_N$ ($T_s=T_N$)
are gradually suppressed with increasing $x$, reaching optimal superconductivity
around $x=0.30$ with $T_c \approx 30$ K.
Here we show that a moderate uniaxial pressure along the $c$-axis
in BaFe$_2$(As$_{0.70}$P$_{0.30}$)$_2$ spontaneously induces a 3D collinear AF order with $T_N=T_s>30$ K, while
only slightly suppresses $T_c$. Although a $\sim$400 MPa pressure
compresses the $c$-axis lattice while
expanding the in-plane lattice and increasing the nearest-neighbor Fe-Fe distance, it
barely changes the average iron-pnictogen height in BaFe$_2$(As$_{0.70}$P$_{0.30}$)$_2$.  Therefore, the
pressure-induced AF order must arise from a strong in-plane
magnetoelastic coupling, suggesting that the 2D nematic phase is a
competing state with superconductivity.
}

~\section{Introduction}

High-transition temperature (high-$T_c$) superconductors such as copper oxides and iron pnictides have plethora of electronic phases that emerge from their three-dimensional (3D) antiferromagnetic (AF) ordered parent compounds \cite{keimer,hosono15,cruz,QHuang08,mgkim,pcdai}.  While the microscopic mechanism of such emergence remains elusive, it is generally believed that the formation of different phases results from a subtle balance among the spin, charge, lattice and orbital degrees of freedom. Therefore, by using an external probe such as pressure to tune this balance, one can manipulate different emergent electronic phases in high-$T_c$ superconductors, and understand their relationship with superconductivity.


In the case of iron pnictides such as BaFe$_2$(As$_{1-x}$P$_{x}$)$_2$ \cite{stewart,Johnston10},
where the parent compound BaFe$_2$As$_2$ has a collinear AF order
below $T_N$ preceded by a tetragonal-to-orthorhombic lattice distortion at $T_s$ with $T_s\geq T_N$ [Figs. 1(a)-1(c)] \cite{QHuang08,mgkim,pcdai}, optimal superconductivity with $T_c\approx 30$ K emerges upon complete
suppression of the 3D AF order and $T_s$ by increasing $x$ to 0.3 \cite{SJiang09,Shibauchi14} or hydrostatic pressure \cite{tomi12,yama10,duncan10}. From transport, magnetic penetration depth, and heat capacity measurements,
a quantum critical point (QCP), where a nonzero temperature phase transition is suppressed continuously
to zero temperature, has been identified near $x\approx 0.3$  \cite{Shibauchi14}. Although initial nuclear magnetic resonance (NMR)
measurements \cite{Nakai10} suggest that the observed QCP arises from the gradual suppression of the collinear AF order near $x=0.3$, neutron diffraction and recent NMR measurements indicate that the AF transition in
BaFe$_2$(As$_{1-x}$P$_{x}$)$_2$ disappears in a weakly first-order fashion
near optimal superconductivity with an avoided magnetic QCP \cite{Allred14,DHu2015,Dioguardi}. On the other hand,
a two-dimensional (2D) nematic phase, whose electronic properties break the in-plane rotational symmetry
but preserve the translational symmetry of the underlying lattice \cite{fradkin}, has been theoretically postulated below $T_s$ \cite{CFang08,Fernandes11,Fernandes14} and experimentally observed in different families of iron pnictides \cite{chu10,matanatar,fisher,Kasahara,xyscience,Rosenthal}.
In particular, elastoresistivity measurements in BaFe$_2$(As$_{1-x}$P$_{x}$)$_2$ reveal a diverging nematic susceptibility
near $x\approx 0.3$, suggesting a nematic origin for the observed QCP \cite{Kuo2016,Analytis}.

Although optimal superconductivity can be
 achieved by a continuous suppression of the finite temperature structural phase transition $T_s$ and magnetic phase transition $T_N$ to zero temperature through increasing P-doping
 in BaFe$_2$(As$_{1-x}$P$_{x}$)$_2$ [Fig. 1(d)] \cite{SJiang09}, it is rather difficult to
study the interplay between the nematic/magnetic order and superconductivity
due to the sensitive P-doping dependence of structural and
magnetic transitions near the optimal superconductivity \cite{Shibauchi14}. Therefore, it is unclear if the nematic phase which is associated with static AF order help or compete with superconductivity.
Although it has been argued theoretically that nematic fluctuations can enhance superconductivity \cite{Metlitski,Lederer}, the relationship between static nematic order and superconductivity is still unclear.

One way to resolve this problem is
to use uniaxial pressure
(strain) as a probe to tune the system toward or away from the optimal superconductivity. In general, $T_c$ in iron based superconductors is related with crystal structural parameters
including the height of As/P anion (pnictogen) to the Fe layer $h_{anion}=h_{As/P}$
[Fig. 1(d)] \cite{zhao08,lee08,kote08,kuro09,xianhui14}.
For BaFe$_2$(As$_{1-x}$P$_{x}$)$_2$, increasing P-doping concentration is linearly associated with decreasing pnictogen
height $h_{As/P}$, and reaches optimal superconductivity with $T_c\approx 30$ K near
$h_{As/P}\approx 1.30$ \AA\ [Fig. 1(d)] \cite{kasa10}. From pressure dependence of thermodynamic measurements, it was found that c-axis uniaxial pressure on BaFe$_2$(As$_{1-x}$P$_{x}$)$_2$ corresponds to an increased P-doping level \cite{tomi12,bohmer12,fisher2012}.
This means that in the P underdoped regime ($x<0.3$), uniaxial pressure along the $c$-axis will increase $T_c$ and suppress $T_N$,
while in the overdoped regime without static AF order ($x\geq 0.3$) it will simply suppress $T_c$.
Since nematic order in iron pnictides occurs below $T_s$ \cite{Fernandes14}, a $c$-axis aligned uniaxial pressure on the
tetragonal structured BaFe$_2$(As$_{0.70}$P$_{0.30}$)$_2$ near the optimal superconductivity
is not expected to induce static collinear AF order, which is associated with the in-plane orthorhombic lattice distortion [Fig. 1(b)].
For BaFe$_2$(As$_{0.70}$P$_{0.30}$)$_2$ near optimal
superconductivity \cite{Allred14,DHu2015}, transport measurements at zero pressure revealed the linear temperature dependence of the resistivity above $T_c$ [$R(T)=R(0)+\alpha T^n$ with $n\approx 1$, where
$R(T)$ is resistivity, $\alpha$ and $n$ are fitting parameters.] [Fig. 1(g)] \cite{Shibauchi14}.
Application of a $c$-axis pressure is expected to push the phase diagram into the overdoped regime, and
should not alter dramatically the linear temperature dependence of the resistivity.

\begin{figure}[t]
\includegraphics[scale=1]{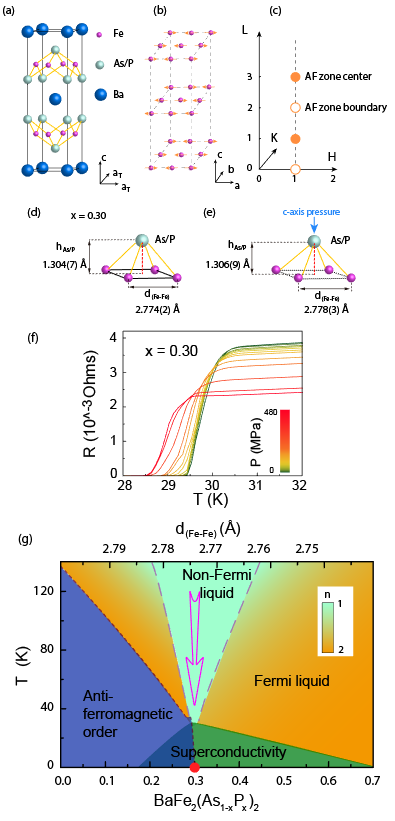}
\caption{{\bf Summary of the electronic phase diagram and uniaxial pressure effect on transport data in BaFe$_2$(As$_{1-x}$P$_x$)$_2$.}
(a) The crystal structure of BaFe$_2$(As$_{1-x}$P$_x$)$_2$. The purple, silvery, and blue balls indicate Fe, As/P, and Ba positions, respectively. (b) The collinear AF structure of BaFe$_2$As$_2$, which requires orthorhombic lattice distortion.
(c) The corresponding reciprocal space, where AF Bragg peaks occur at $(1,0,L)$ with $L=1,3,\cdots$.
(d) Schematic diagrams of the FeAs tetrahedron, showing the (As, P) height to iron plane is
$h_{As/P}=1.304(7)$ \AA\ and the nearest in plabe Fe-Fe atom distance is $2.774(2)$ \AA, respectively,
for BaFe$_2$(As$_{0.70}$P$_{0.30}$)$_2$ in 300K. (e) By applying an uniaxial pressure along the c-axis, pnictogen height is $h_{As/P}=1.306(9)$ \AA\, and the in-plane Fe-Fe atom distance increases to $2.778(3)$ \AA\ correspondingly .
(f) Temperature dependence of in-plane resistance for pressures up-to 480 MPa along the $c$-axis of
BaFe$_2$(As$_{0.70}$P$_{0.30}$)$_2$ crystal.
(g) The electronic phase diagram of BaFe$_2$(As$_{1-x}$P$_x$)$_2$,
where AF, superconductivity, Fermi liquid, Non-Fermi liquid, and a QCP (red dot) are marked.
The yellow and green colors represent the values of
$n$ in temperature dependence of the resistivity exponent \cite{Shibauchi14}. Arrow marks the position in the phase diagram of the compound measured in this work.
}
\end{figure}

\begin{figure}[t]
\includegraphics[scale=.8]{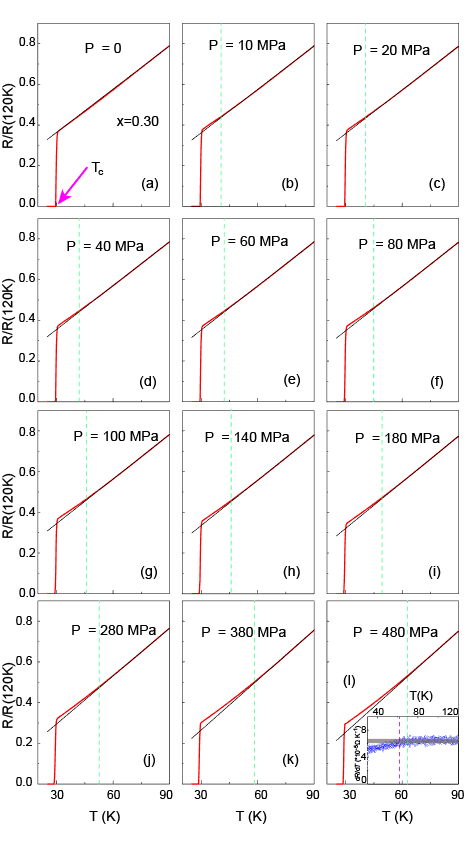}
\caption{
{\bf Temperature dependence of normalized in-plane resistance under different $c$-axis pressure.}
Temperature dependence of normalized in-plane resistance for different pressures along
the $c$-axis of BaFe$_2$(As$_{0.70}$P$_{0.30}$)$_2$ crystal using automated uniaxial pressure
device described in Ref. \cite{Tam17}. (a) $P=0$. The arrow indicates $T_c$ at $\rho=0$.
(b) $P=10$ MPa. (c) $P=20$ MPa. (d) $P=40$ MPa. (e) $P=60$ MPa. (f) $P=80$ MPa.
(g) $P=100$ MPa. (h) $P=140$ MPa. (i) $P=180$ MPa. (j) $P=280$ MPa. (k) $P=380$ MPa.
(l) $P=480$ MPa. The inset shows temperature derivative of the
in-plane resistance, which reveal more clearly the deviation from
linear temperature dependence. The black lines are the linear fits for high temperature data (80 K-120 K) and
the vertical dashed green lines mark the temperature where the observed resistance deviates
from the linear temperature dependence.
 }
 \end{figure}

Surprisingly, we find that a $c$-axis aligned uniaxial pressure on BaFe$_2$(As$_{0.70}$P$_{0.30}$)$_2$
not only suppresses $T_c$ [Fig. 1(f)], but also induces resistivity
deviation from the zero pressure linear temperature dependence [Fig. 2].
Figure 3 shows uniaxial pressure dependence of $T_c$, suggesting an approximate linear relationship
between the applied pressure and reduction in $T_c$ below $\sim$280 MPa.
Upon further increasing uniaxial pressure, the reduction in $T_c$ reduces, while the temperature where
resistivity deviates from the linear temperature dependence increases (Fig. 3).
Our neutron diffraction experiments on BaFe$_2$(As$_{0.70}$P$_{0.30}$)$_2$
 reveal that the deviation in resistivity from the linear temperature dependence
in Figs. 2 and 3 arise from a
pressure-induced collinear AF order with $T_N> T_c$ (Figs. 4 and 5).
This is difficult to understand
within the conventional Fermi surface nesting picture \cite{Hirschfeld2011},
where a $c$-axis pressure is expected to induce pnictogen height reduction,
promote electron itinerancy, and drive the system to the P-overdoped side with reduced $T_c$ and no static AF order.
Since the collinear AF order in iron pnictides has to be associated with the orthorhombic
distortion of the underlying lattice [Fig. 1(b)] \cite{cruz,QHuang08,mgkim,pcdai},
our results provide the compelling evidence that a $c$-axis pressure actually
induces an in-plane $C_4$ symmetry breaking field in
the underlying lattice and drives the system into a nematic ordered phase.
To understand this phenomenon, we carried out neutron diffraction experiments on single crystals with
zero and $\sim$400 MPa $c$-axis pressure.  Our Rietveld analysis reveals that the dominant effect of a $c$-axis pressure is
to suppress the $c$-axis and expand the in-plane lattice constants, and has limited effect on the average (As,P) pnictogen
height $h_{As/P}$ [Fig. 1(d) and 1(e), and Table I]. This is true even though As and P atoms
occupy distinct positions in BaFe$_2$(As$_{1-x}$P$_{x}$)$_2$ \cite{rotter10}.
Based on the refined crystal structures at zero and finite pressure,
we use a combination of density functional theory and dynamical mean field theory (DFT+DMFT) \cite{ DMFT-RMP2006}
to study the effect of a $c$-axis
pressure on the magnetic property of BaFe$_2$(As$_{0.70}$P$_{0.30}$)$_2$, and find that strong magnetoelastic coupling with the in-plane lattice is the driving force for the c-axis pressure induced long-range AF order.
As the pressure-induced nematic order with $T_N,T_s > 30 $ K increases with increasing pressure (Figs. 2-5), we conclude that the nematic order is competing with superconductivity and therefore
needs to be taken into account to understand the electronic properties of iron pnictides.

~\section{Results}

~\subsection{Resistivity measurements}

We chose to study the effect of a $c$-axis aligned uniaxial pressure on BaFe$_2$(As$_{0.70}$P$_{0.30}$)$_2$ because P-substitution does not introduce significant
disorder \cite{analy10,beek10,klin10} and is expected to reduce the average pnictogen
height linearly similar to $c$-axis applied pressure \cite{kasa10}.
Previous transport measurements on BaFe$_2$(As$_{1-x}$P$_x$)$_2$ single crystals reveal
that both the AF order and superconductivity are sensitive to the applied pressure \cite{fisher2012}.
Similar to other electron-doped iron pnictides \cite{dhital,Dhital14,YSong13,man,XYLu15}, in-plane uniaxial pressure
that breaks the $C_4$ symmetry of the
underlying lattice enhances
the AF order and suppresses superconductivity in underdoped BaFe$_2$(As$_{1-x}$P$_x$)$_2$.
For comparison, a $c$-axis aligned uniaxial pressure increases
$T_c$ and suppresses $T_N$ in the underdoped regime, and is expected to only suppress $T_c$ in the optimal/overdoped regime [Fig. 1(g)] \cite{bohmer12,fisher2012}. In addition, a $c$-axis pressure is also expected to expand the in-plane lattice parameter, akin
to negative in-plane pressure.

We first describe $c$-axis pressure dependence of the resistivity on BaFe$_2$(As$_{0.70}$P$_{0.30}$)$_2$ \cite{SI}.
Single crystals of BaFe$_2$(As$_{0.70}$P$_{0.30}$)$_2$ were prepared using the self-flux method \cite{DHu2015}. Transport measurements were carried out using a commercial physical property measurement system (PPMS) with
the standard four-probe method. Figure 1(f) shows pressure dependence of the resistivity near $T_c$. With increasing pressure from
0 to 480 MPa using a custom designed uniaxial pressure device \cite{Tam17}, we see a systematic reduction in $T_c$ from 29.5 K to 28.5 K.
Figure 2 summarizes temperature dependence of the resistivity as a function of increasing pressure. At zero pressure,
we find linear temperature dependence of the resistivity above $T_c$ and below 90 K, confirming previous work \cite{Shibauchi14}. With increasing pressure from 0 to 480 MPa, temperature dependence of the resistivity
systematically deviates from the linear behavior as marked by the vertical green dashed lines in Fig. 2.
This is reminiscent of the situation in
BaFe$_2$(As$_{0.71}$P$_{0.29}$)$_2$ at ambient condition, where the system orders
antiferromagnetically at a temperature above $T_c$ \cite{DHu2015}. Assuming that the deviation from the
linear temperature dependence is due to pressure-induced static AF order at $T_N$, we show in Fig. 3 pressure dependence of
$T_c$ and $T_N$. Simple linear fits to the data yield a reduction in $T_c$ of $3.0\pm 0.2$ K/GPa, and an increase of
$T_N$ of $48\pm 2$ K/GPa if we ignore the $T_N=0$ at zero pressure point (Fig. 3).
From neutron diffraction experiments described below, we know that $c$-axis
pressure induced AF ordered moment in BaFe$_2$(As$_{0.70}$P$_{0.30}$)$_2$ is vanishingly small compared with that of the parent compound BaFe$_2$As$_2$. Therefore, we do not expect
a $c$-axis pressure will be able to induce observable magnetic ordered moment change
in BaFe$_2$As$_2$.

~\subsection{Neutron scattering results}

To confirm the results of transport measurements, we have carried out neutron scattering experiments using
the MIRA triple-axis spectrometer at Maier-Leibnitz Zentrum, Garching, Germany \cite{robert15}, HB-1A triple-axis spectrometer and HB-3A
four circle diffractometer at High Flux Isotope Reactor, Oak Ridge National Laboratory.
Two single crystals of BaFe$_2$(As$_{0.70}$P$_{0.30}$)$_2$, labeled A and B,
 were clamped between two Al plates using a pressure cell \cite{SI}.
We define the wave vector \textbf{Q} in three-dimensional reciprocal space in \AA$^{-1}$ as ${\bf Q}=H {\bf a^\ast}+K{\bf b^\ast}+L{\bf c^\ast}$, where $H$, $K$, and $L$ are Miller indices and ${\bf a^\ast}=\hat{{\bf a}}2\pi/a, {\bf b^\ast}=\hat{{\bf b}} 2\pi/b, {\bf c^\ast}=\hat{{\bf c}}2\pi/c$ are reciprocal lattice units (r.l.u.) [Fig. 1(c)].
In the low-temperature AF orthorhombic phase of BaFe$_2$(As$_{1-x}$P$_x$)$_2$, $a\approx 5.57$ \AA, $b\approx 5.55$ \AA, and $c\approx 12.80$ \AA\  \cite{Allred14}. In this notation,
we expect magnetic Bragg peaks of
the collinear AF structure to occur at $(1,0,L)$ with $L=1,3,5$ positions [Fig. 1(c)] \cite{QHuang08,mgkim,pcdai}. Near optimal
superconductivity, the sample is a tetragonal paramagnet with $a=b$. We first carried out triple-axis measurements to determine the effect of
uniaxial pressure on magnetic order. In order to determine the effect of $c$-axis uniaxial pressure on crystalline lattice,
we measured many nuclear Bragg peaks in sample A after the triple axis measurements under pressure at room temperature.
We then took another crystal from the same batch and put it in the identical
pressure cell, and carried out the same diffraction measurement without pressure at room temperature.
These two measurements allowed us to determine the effect of $c$-axis pressure on the
atomic positions and lattice parameters of BaFe$_2$(As$_{0.70}$P$_{0.30}$)$_2$ \cite{SI}.

\begin{figure}[t]
\includegraphics[scale=.6]{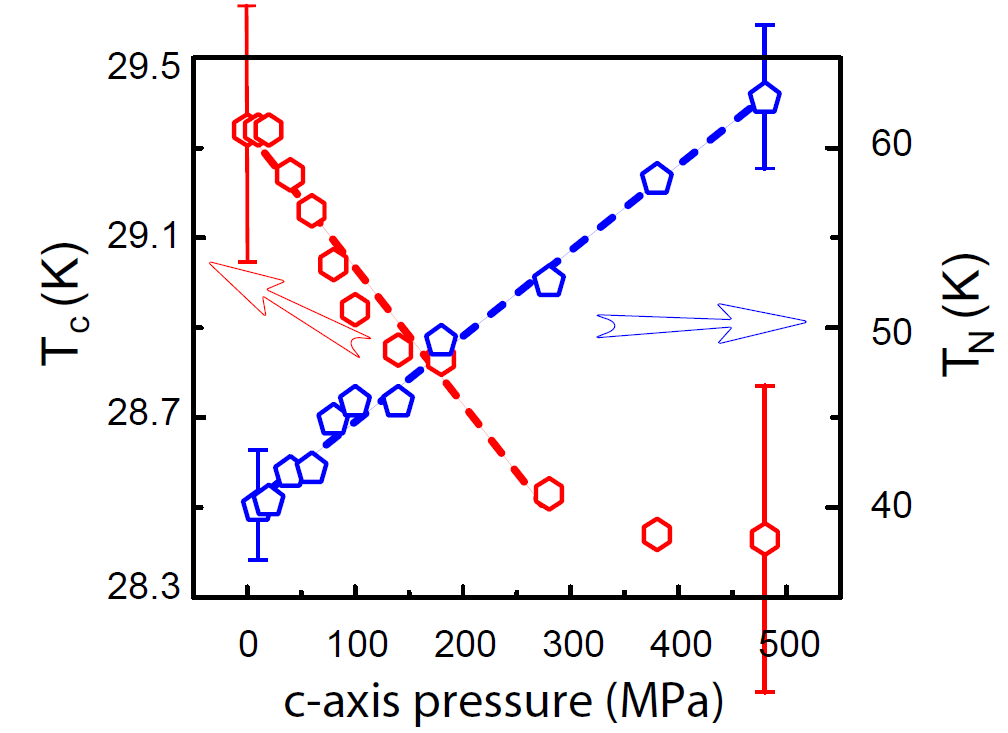}
\caption{
{\bf The $c$-axis pressure dependence of $T_c$ and $T_N$ as determined from transport measurement.}
The $c$-axis uniaxial pressure dependence of $T_c$ at $\rho=0$ and the deviation temperatures
from the linear temperature resistance. The error bars are the superconducting transition widths with $P_c$ = 0 and 480 MPa.
The error bars for $T_N$ are estimated by using temperature derivative of the resistance data.
The red and blue dashed lines are linear fits to the data.
 }
\end{figure}

Figure 4 summarizes the triple-axis neutron scattering results at zero and finite $c$-axis uniaxial pressure on sample A.
At zero pressure, the sample has no static AF order
at $T=29$ K and 6 K as shown in wave vector scans along the $[H,0,3]$ direction
through the magnetic Bragg peak position $(1,0,3)$ [Figs. 4(a) and 4(b)]. Upon application of a
$c$-axis aligned uniaxial pressure $P_a$,
we find temperature dependent peaks at $(1,0,3)$ [Figs. 4(a)-4(c)].
The solid lines in Figs. 4(a)-4(c) are Gaussian fits to the data, which give the
static spin-spin correlation lengths of $220\pm 30$ \AA,
indicating pressure induced long-range magnetic order.
To determine the spin arrangements in the pressure induced magnetic phase,
we measured magnetic Bragg peaks at different $L$ positions along the $[H,0]$ direction.
Figure 4(d) shows wave vector scans along the $[H,0]$ direction at $L=1,2,3$. For collinear AF order in
underdoped BaFe$_2$(As$_{1-x}$P$_x$)$_2$,
we expect magnetic Bragg peaks at $L=1,3$ and no magnetic scattering $L=2$ \cite{QHuang08,mgkim,pcdai}.
Figure 4(d) shows that the magnetic scattering around $(1,0,3)$ is about 3.1 times of that at $(1,0,1)$, and no magnetic signal
at $(1,0,2)$. These results are consistent with the collinear AF order in underdoped BaFe$_2$(As$_{1-x}$P$_x$)$_2$ \cite{DHu2015}.
By normalizing the magnetic Bragg peak intensity with a weak nuclear Bragg peak, we estimate that the pressure-induced magnetic
ordered moment is about 0.1 $\mu_B$/Fe.
Figure 4(e) compares temperature dependence of the $(1,0,3)$ scattering at zero ($P=0$) and
finite ($P=P_a$) $c$-axis pressure. While the scattering has no temperature
dependence at zero pressure consistent with no magnetic order,
its behavior at $P_a$ is similar to those of
underdoped BaFe$_2$(As$_{1-x}$P$_x$)$_2$ with $T_N\approx 55$ K, where
the reduction of magnetic scattering below $T_c$ is due to the competition of
 superconductivity with static AF order \cite{DHu2015}.

Assuming that the temperature where resistivity deviates from the
linear temperature dependence in $c$-axis pressured sample signals the start of
static AF order, we compare $T_N$ measured by neutron scattering in Fig. 4 with Fig. 3 determined from
transport measurements and find $P_a\approx 400$ MPa. The estimated $P_a$ is consistent
with the observed reduction in $T_c$ as seen in magnetic scattering intensity
reduction in Fig. 4(e), although $T_c$ of the sample can only be approximately determined by such measurement.
To accurately determine the effect of $c$-axis uniaxial pressure on the pnictogen
height $h_{As/P}$ and lattice parameters of
BaFe$_2$(As$_{0.70}$P$_{0.30}$)$_2$, we measured 30 Bragg peaks with $P=P_a$
and repeated the same measurement without pressure at room temperature using HB-3A four
circle diffractometer \cite{SI}.  The Rietveld analysis of the single crystal diffraction
data reveal that the major effect of a $P\approx 400$ MPa pressure is to suppress the $c$-axis and expand the in-plane lattice
parameters without affecting
much the average  pnictogen height $h_{As/P}$ (Table I).

\begin{table*}
\caption{\label{tab:table1}
{\bf Results of single crystal diffraction experiments on sample A with $P=P_a$ and $P=0$.}
The refined lattice constants, Fe-Fe distance, and $h_{As/P}$ of the $x=0.30$ sample at 295 K.
}
\begin{ruledtabular}
\begin{tabular}{cccccc}
$ $    & $a=b$ & $c$  & $d_{(Fe-Fe)}$  & $z_{As/P}$   & $h_{As/P}$    \\
$P=0$    & $5.548(4)$ & $12.817(7)$  & $2.774(2)$  & $0.3517(5)$    & $1.304(7)$  \\
$P=P_a$ & $5.556(6)$  & $12.778(7)$ & $2.778(3)$   & $ 0.3522(7)$   & $1.306(9)$     \\
\end{tabular}
\end{ruledtabular}
\end{table*}

To determine the uniaxial pressure dependence of the magnetic order, we carried out
neutron scattering measurements on sample B under pressure $P=P_b<P_a$.
Figure 5(a) shows temperature dependence of the magnetic scattering at
$(1,0,3)$ at zero and $P=P_b$. Similar to sample A, we find that uniaxial pressure induced
magnetic order first appears below $T_N\approx 42$ K, competes with superconductivity below $T_c$,
and is completely suppressed below 10 K. The lower $T_N$ in sample B suggests $P_b\approx 40$ MPa based
on transport measurements in Fig. 3. Figures 5(b), 5(c), and 5(d) compare zero and finite pressure ($P=P_b$) wave
vector scans around $(1,0,3)$ at different temperatures. From these scans, we estimate that pressure
induced AF order has in-plane and $c$-axis spin-spin correlation lengths of about 120 \AA\ and 160 \AA, respectively.
These results are consistent with those described in Fig. 4, revealing that a $c$-axis aligned uniaxial pressure
spontaneously induces the collinear AF order with $T_N>30$ K.

~\subsection{DFT+DMFT calculation}

Our observation suggests that the magnetic properties of BaFe$_2$(As$_{0.70}$P$_{0.30}$)$_2$
is very sensitive to small structure changes induced by the $c$-axis pressure at the optimal P-doping,
signaling a large magnetoelastic coupling.
Earlier DFT calculations \cite{ZPY08} of the LaFeAsO parent compound revealed a strong dependence of
the ordered magnetic moment on the As height.
However, DFT calculations consistently overestimate the Fe ordered magnetic moment \cite{ZPY11_2}.
Although DFT+DMFT calculations can accurately reproduce the experimental ordered magnetic moments of Fe atoms
in a large numbers of iron-based compounds
with the same Hubbard $U$ and Hund's coupling $J$ as used in the current study \cite{ZPY11_2},
the dependence of the Fe magnetic moment on the As height has not been carefully studied by DFT+DMFT calculations.

To determine the magnitude of the magnetoelastic coupling and understand the unusual behavior under $c$-axis pressure,
we use a DFT+DMFT theory to study BaFe$_2$(As$_{0.70}$P$_{0.30}$)$_2$ \cite{DMFT-RMP2006}.
In the DFT+DMFT calculations, the electronic charge was computed self-consistently on DFT+DMFT density matrix.
The quantum impurity problem was solved by the continuous time quantum Monte Carlo method \cite{Haule-QMC,Werner},
at a temperature of 72.5 K if not otherwise specified,
 and with a Hubbard $U=5.0$ eV and Hund's rule coupling $J=0.7$ eV in the AF state \cite{ZPY11,ZPY11_2,Tam17}.
The experimental lattice constants of optimal P-doped BaFe$_2$(As$_{0.7}$P$_{0.3}$)$_2$, $a=b=5.5406$ \AA\ and $c=12.761$ \AA\ \cite{Johnston10} are used in the calculations
while the internal As/P position $z_{As/P}$ [where $h_{As/P}=(z_{As/P}-0.25)c$] is varied from 0.349 to 0.347
in order to determine the (theoretical) critical As/P position/height.
In the other set of calculations, the effect of $c$-axis pressure on the lattice is taken into account by expanding
the in-plane lattice constant $a$ by 1\% and varying the internal As/P position in the same range whereas the $c$-lattice constant
is fixed to the experimental value to simplify the computation since the As/P position/height and in-plane lattice constant are the dominating factors in determining the magnetic properties.
By varying both the As/P height and the in-plane lattice constant, we are able to compute the magnitude of the in-plane and $c$-axis magnetoelastic coupling constants.

\begin{table*}
\caption{\label{tab:table2} The DFT+DMFT calculated Fe ordered magnetic moment in the collinear AF state as a function of As/P height $z_{As/P}$ at the experimental and 1\% expanded in-plane lattice constant while keep the $c$-lattice constant unchanged.
}
\begin{ruledtabular}
\begin{tabular}{cccccc}
$z_{As/P}$        & 0.349 & 0.3485 & 0.348 & 0.3475 & 0.347 \\
$\Delta a/a=0$    & 0.138 & 0.065  & $<0.002$  & $<0.001$   & $<0.001$     \\
$\Delta a/a=0.01$ & 0.217  & 0.143 & 0.049   & 0.002       & $<0.001$  \\
\end{tabular}
\end{ruledtabular}
\end{table*}

We summarize our DFT+DMFT results in Table II. Keeping the lattice constant $a$ and $c$ unchanged and varying
only the As/P height $z_{As/P}$ (second row in Table II),
the Fe ordered magnetic moment is 0.138 $\mu_B$/Fe at $z_{As/P}=0.349$, and quickly decreases to 0.065 $\mu_B$/Fe at $z_{As,P} =0.3485$, and less than 0.002 $\mu_B$/Fe at $z_{As,P}=0.348$.
The Fe ordered magnetic moment changes at a rate of ~11 $\mu_B/$\AA\ with the $h_{As/P}$, much larger than the corresponding
value of $\sim3.7\ \mu_B/$\AA\ in LaFeAsO \cite{ZPY08}.
Therefore, the DFT+DMFT magnetoelastic coupling strength in optimal P-doped BaFe$_2$(As$_{1-x}$P$_x$)$_2$ is about 3 times stronger than the DFT magnetoelastic coupling strength in LaFeAsO,
suggesting correlation-enhanced spin-phonon coupling \cite{ZPY13,Mandal14,Gerber17}.

Keeping the (As,P) height unchanged at $z_{As/P}=0.348$ as a function of increasing pressure,
as seen experimentally in BaFe$_2$(As$_{0.70}$P$_{0.30}$)$_2$ (Table I), increasing the in-plane Fe-Fe spacing by 1\% increases the Fe ordered moment to 0.049 $\mu_B$/Fe, a change rate of ~1.2 $\mu_B/$\AA, an order of magnitude smaller than the change rate with respect to
the (As,P) height. Therefore, the (As,P) height is the dominating factor in the magnetic properties of the compound.
This is further supported by the corresponding calculations at $z_{As/P}=0.3475$ and 0.347 with experimental and 1\% expanded in-plane lattice constant (Table II). With slightly reduced $z_{As/P}$, expanding the in-plane Fe-Fe spacing
by 1\% barely increases the Fe ordered magnetic moment,
in strong contrast to the results at $z_{As/P}=0.348$. Therefore, the $z_{As/P}=0.348$ is a critical (As,P) height
where the Fe ordered magnetic moment is very sensitive to the in-plane Fe-Fe distance, similar to a previous observation that the Fe ordered magnetic moment shows a large enhancement
under in-plane uniaxial pressure near a optimal superconductivity in BaFe$_{2-x}T_x$As$_2$ ($T=$Co,Ni) \cite{Tam17}.

Finally, we adopt the experimental results and carry out additional DFT+DMFT calculations in the AF state
by expanding the in-plane lattice constant $a$ by 0.5\% (from 3.9178 \AA\ to 3.9374 \AA ) and compressing the $c$-axis lattice constant by 1\% (from 12.761 \AA\  to 12.6347 \AA)
while the internal As/P position $z_{As/P}$ is changed from 0.348 to 0.349 in order to keep the As/P height unchanged.
The calculations are done at a temperature of 29 K, below the experimental N$\rm \acute{e}$el temperature.
We find that, without the $c$-axis pressure, the ordered Fe moment
is still less than 0.01 $\mu_B$, implying no long-range AF magnetic order. However, the long-range AF magnetic order emerges with an ordered Fe moment of 0.10 $\mu_B$ upon applying the aforementioned $c$-axis pressure.
These findings are in good agreement with
our experimental obvervation.

~\section{Discussion}

Our transport, neutron scattering experiments, and DFT+DMFT calculations on BaFe$_2$(As$_{0.70}$P$_{0.30}$)$_2$
reveal that magnetism in this material is particularly sensitive to a $c$-axis aligned
uniaxial pressure. In previous transport, NMR, magnetic penetration depth, and heat capacity measurements,
a QCP has been found at BaFe$_2$(As$_{0.70}$P$_{0.30}$)$_2$ \cite{Shibauchi14}. In particular,
differential elastoresistance measurements on BaFe$_2$(As$_{0.70}$P$_{0.30}$)$_2$ indicate a
diverging nematic susceptibility at $T=0$, suggesting that the QCP is
nematic in origin \cite{Kuo2016}. Since nematic fluctuations may enhance superconductivity \cite{Metlitski,Lederer},
it would be important to sort out the relationship between the static nematic phase and superconductivity.

In principle, nematic order of iron pnicitdes is an electronic anisotropic property of the 2D FeAs plane and should couple linearly
to anisotropic strain (or pressure) within the FeAs plane in the limit of infinitesimal strains \cite{Fernandes14}.
By measuring the rate of change of resistivity anisotropy with respect to the in-plane anisotropic strain, one can determine the nematic
susceptibility \cite{Kuo2016}.
By contrast, uniaxial pressure along the $c$-axis, which does not break the in-plane crystalline lattice symmetry, exhibits nonlinear coupling with the nematic order parameter \cite{fujii2018}.
Our surprising discovery that a $c$-axis aligned strain can actually induce
AF order with in-plane symmetry-breaking field suggests that nematic order can also couple to the $c$-axis pressure
via pressure-induced in-plane Fe-Fe distance expansion (Table I). Although a $\sim$400 MPa pressure along the $c$-axis suppresses the
$c$-axis and expands the in-plane lattice parameters of
BaFe$_2$(As$_{0.70}$P$_{0.30}$)$_2$, it has negligible effect on the (As,P) height $h_{As/P}$ [Fig. 1(d) and 1(e), and Table I].
As demonstrated by our DFT+DMFT calculation, this means that BaFe$_2$(As$_{0.70}$P$_{0.30}$)$_2$ is at critical doping regime with a critical As/P height
where the Fe ordered moment and associated nematic order depend sensitively on the in-plane Fe-Fe distance.
In the underdoped region, the As/P height is larger than the critical value of $h_{As/P}\approx 1.30$ \AA\ and
the long-range AF order already exists at ambient pressure. Applying a $c$-axis pressure does not change the AF order as dramatically as it does
at the optimal doping. On the other hand, in the overdoped region, where the As/P height is below the critical value,
a $c$-axis pressure of similar magnitude should be unable to induce the long-range AF order as shown in Table II.
Further pressure dependent transport and neutron scattering experiments as a function of P-doping will help us to better understand the quantum critical behavior in iron-based superconductors.

The $c$-axis pressure-induced AF order we observed in BaFe$_2$(As$_{0.70}$P$_{0.30}$)$_2$ is in someways reminiscent
of the effect of biaxial strain on the phase transitions in Ca(Fe$_{1-x}$Co$_x$)$_2$As$_2$  \cite{bohmer17,Ca122-Boehmer2}.  Similar to $c$-axis uniaxial pressure, biaxial in-plane strain achieved by using the differential thermal
expansion between the samples and a rigid substrate affects the $c/a$ ratio of the
tetragonal samples but does not break the tetragonal symmetry \cite{bohmer17}.  Instead of decreasing the $c/a$ ratio by applying
a $c$-axis pressure, biaxial in-plane strain in Ca(Fe$_{1-x}$Co$_x$)$_2$As$_2$ increases the $c/a$ ratio and shifts
the phase diagram of Ca(Fe$_{1-x}$Co$_x$)$_2$As$_2$ to higher $x$, without changing the maximum $T_c$ and magnetic phase
transition temperature \cite{bohmer17}.  Here, the quantum phase transition between the  superconducting tetragonal and AF orthorhombic phases is first-order, and the resulting phase separation into these two phases with different in-plane lattice parameters allows the material to respond to biaxial strain in a continuous fashion \cite{Ca122-Boehmer2}.
For comparison with BaFe$_2$(As$_{0.70}$P$_{0.30}$)$_2$, it would be interesting to experimentally determine the biaxial pressure dependence of the iron pnictogen height and Fe-Fe distance, and use DFT+DMFT
calculation to see if the experimental phase diagram can be reproduced.

In summary, our neutron scattering results reveal that a moderate uniaxial pressure along the $c$-axis
in BaFe$_2$(As$_{0.7}$P$_{0.3}$)$_2$ superconductor can spontaneously induce a three-dimensional stripe AF order with $T_N>30$ K, while only slightly suppresses $T_c$. These results indicate a strong magnetoelastic coupling near optimal superconductivity, suggesting
an uniaxial pressure induced nematic phase and stripe AF order competing with superconductivity
in BaFe$_2$(As$_{1-x}$P$_x$)$_2$.

\begin{flushleft}

{\bf Data availability.}
The data that support the findings of this study are available
from the corresponding authors on request.

\end{flushleft}

\begin{figure}[t]
\includegraphics[scale=.8]{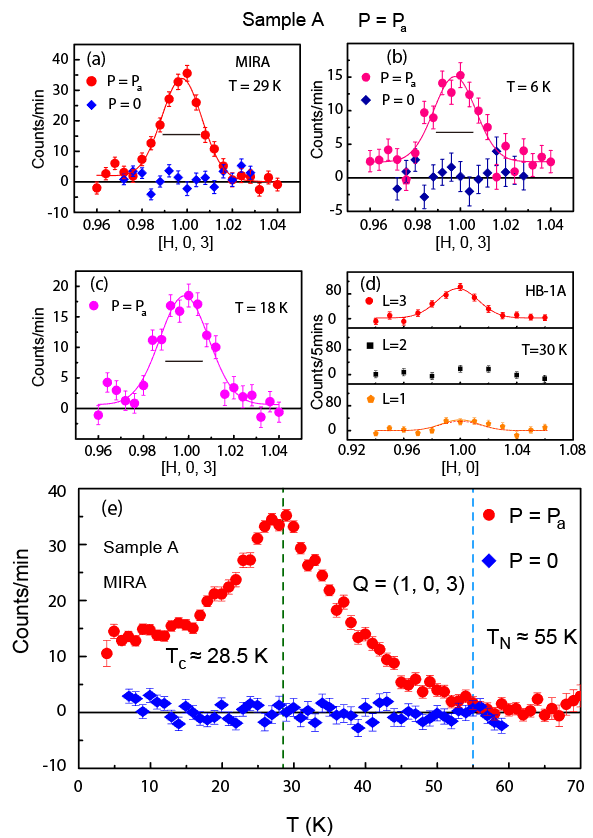}
\caption{{\bf Summary of neutron scattering results on sample A}.
The c-axis pressure-induced magnetic signal in sample A carried out at MIRA and HB-1A.
(a-c) Wave vector scans through the $(1,0,3)$ magnetic Bragg peak position at different
temperatures with zero and finite pressure. The solid lines are Gaussian fits to the data,
and the horizontal bars are instrumental resolution. (d) Wave vector scans along the
$[H,0,L]$ direction with $L=1,2,3$ at $T=30$ K and finite pressure.
(e) Temperature dependence of the magnetic scattering at $\textbf{Q}_{AF}=(1,0,3)$ for $P = P_a$
and P = 0. All shown data are background corrected.
}
\end{figure}

\begin{figure}
\includegraphics[scale=.65]{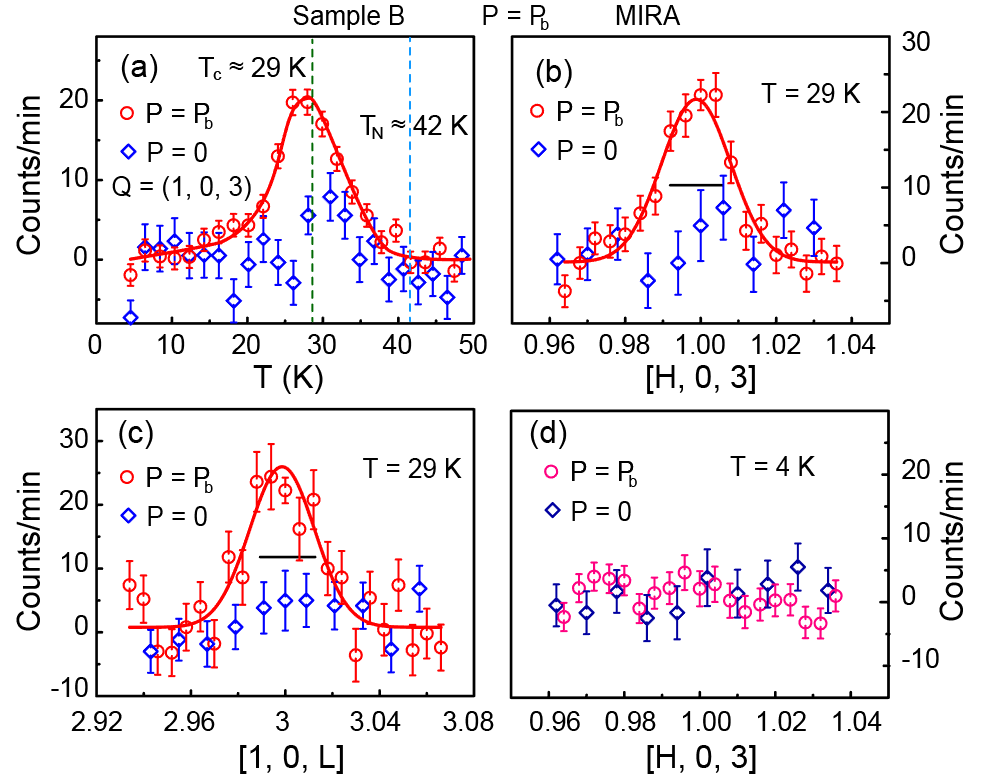}
\caption{{\bf Summary of neutron scattering results on sample B}.
Wave vector and temperature dependence of the uniaxial pressure induced magnetic order
for sample B. (a) Temperature dependence of the magnetic scattering at $(1,0,3)$ with and without pressure.
(b-d) Wave vector scans along the $[H,0,3]$ direction with and without uniaxial pressure at different
temperatures.
The absence of magnetic scattering at low-temperature ($4$ K) suggests that superconductivity can completely
suppress the pressure induced magnetic order. The solid lines in (b) and (c)
are Gaussian fits to the data, and the horizontal bars are instrumental resolution. The solid line in (a)
is a guide to the eye. The blue and green vertical dashed lines in (a)
mark $T_N$ and $T_c$, respectively.
}
\label{fig5}
\end{figure}

\begin{flushleft}

{\bf Acknowledgments}

The neutron scattering work at Rice is supported by the
U.S. NSF-DMR-1700081 (P.D.). A part of the material synthesis work at Rice is
supported by the Robert A. Welch Foundation Grant No. C-1839 (P.D.).
Z.P.Y was supported by the NSFC (Grant No. 11674030), the Fundamental Research Funds for the Central Universities (Grant No.310421113) and the National Key Research and Development Program of China through Contract No. 2016YFA0302300.
The calculations used high performance clusters at the National Supercomputer Center in Guangzhou.
The work at IOP is supported by the "Strategic Priority Research Program (B)" of Chinese Academy of Sciences (XDB07020300), MOST (2012CB821400, 2011CBA00110,2015CB921302,2016YFA0300502), and NSFC (No. 11374011, 11374346, 91221303,11421092,11574359).This research used resources at the High Flux Isotope Reactor, a DOE Office of Science User Facility operated by the Oak Ridge National Laboratory.

\end{flushleft}

\begin{flushleft}
{\bf Author contributions}

Most of the single crystal growth and neutron scattering experiments were carried out by
D.H. with assistance from W.Y.W., W.L.Z., Y.W., D.L.G., Y.S., R.G., B.P.,
H.B.C., W.T., B.R.. Transport measurements are carried out at Rice with help from D.W.T., P.P.Z.
DFT+DMFT calculations are carried by Z.P.Y.
P.D. provided the overall lead of the project.
The paper was written by H.D., Z.P.Y., and P.D.
All authors provided comments.
\end{flushleft}
\begin{flushleft}
{\bf Additional information}

Competing Interests: the authors declare no competing interests.

Correspondence and
requests for materials should be addressed to
D.H. (dinghu@rice.edu), Z.P.Y. (yinzhiping@bnu.edu.cn) or P.D. (e-mail: pdai@rice.edu)
\end{flushleft}


\begin{thebibliography}{}

\bibitem{keimer} Keimer, B., Kivelson, S.A., Norman, M.R., Uchida, S., \& Zaanen, J. From quantum matter to high-temperature superconductivity in copper oxides. Nature {\bf 518}, 179-186 (2015).

\bibitem{hosono15} Hosono, H. ,\& Kuroki, K. Iron-based superconductors: Current status of materials and pairing mechanism. Physica C: Superconductivity and its Applications {\bf 514}, 399-422 (2015).

\bibitem{cruz} de la Cruz, C. {\it et al}. Magnetic order close to superconductivity in the iron-based layered LaO$_{1-x}$F$_x$FeAs systems. Nature {\bf 453}, 899 (2008).

\bibitem{QHuang08} Huang, Q. {\it et al.} Neutron-diffraction measurements of magnetic order and a structural transition in the parent BaFe$_2$As$_2$ compound of FeAs-based high-temperature superconductors. Phys. Rev. Lett. {\bf 101}, 257003 (2008).

\bibitem{mgkim} Kim, M. G. {\it et al.} Character of the structural and magnetic phase transitions in the parent and electron-doped BaFe$_2$As$_2$ compounds. Phys. Rev. B {\bf 83}, 134522 (2011).

\bibitem{pcdai} Dai, P. Antiferromagnetic order and spin dynamics in iron-based superconductors. Rev. Mod. Phys.  {\bf 87}, 855-896 (2015).

\bibitem{stewart} Stewart, G.R., Superconductivity in iron compounds. Rev. Mod. Phys. {\bf 83}, 1589-1652 (2011).

\bibitem{Johnston10} Johnston, D. C., The puzzle of high temperature superconductivity in layered iron pnictides and chalcogenides. Adv. in Phys. {\bf 59}, 803-1061 (2010).

\bibitem{SJiang09} Jiang, S. {\it et al.} Superconductivity up to 30 K in the vicinity of the quantum critical point in BaFe$_2$(As$_{1-x}$P$_x$)$_2$. J. Phys. Condens. Matter. {\bf 21}, 382203 (2009).

\bibitem{Shibauchi14} Shibauchi, T., Carrington, A., $\&$ Matsuda, Y. A Quantum Critical Point Lying Beneath the Superconducting Dome in Iron Pnictides. Annual Review of Condensed Matter Physics {\bf 5}, 113-135 (2014).

\bibitem{tomi12} Tomi$\rm \acute{c}$, M., Valent, R., $\&$ Jeschke, H.O. Uniaxial versus hydrostatic pressure-induced phase transitions in CaFe$_2$As$_2$ and BaFe$_2$As$_2$. Phys. Rev. B {\bf 85} 094105 (2012).

\bibitem{yama10} Yamazaki, T. {\it et al.}, Appearance of pressure-induced superconductivity in BaFe$_2$As$_2$ under hydrostatic conditions and its extremely high sensitivity to uniaxial stress. Phys. Rev. B {\bf 81}, 224511 (2010).

\bibitem{duncan10} Duncan, W. J. {\it et al.}, High pressure study of BaFe$_2$As$_2$--the role of hydrostaticity and uniaxial stress. J. Phys. Condens. Matter. {\bf 22} 052201 (2010).

\bibitem{Nakai10} Nakai, Y. {\it et al.}, Unconventional superconductivity and antiferromagnetic quantum critical behavior in the isovalent-doped BaFe$_2$(As$_{1-x}$P$_x$)$_2$. Phys. Rev. Lett. {\bf 105}, 107003 (2010).

\bibitem{Allred14} Allred, J.M. {\it et al.}, Coincident structural and magnetic order in BaFe$_2$(As$_{1-x}$P$_x$)$_2$ revealed by high-resolution neutron diffraction. Phys. Rev. B {\bf 90}, 104513 (2014).

\bibitem{DHu2015} Hu, D. {\it et al.}, Structural and Magnetic Phase Transitions
near Optimal Superconductivity in BaFe$_2$(As$_{1-x}$P$_x$)$_2$. Phys. Rev. Lett. {\bf 114},157002(2015).

\bibitem{Dioguardi} Dioguardi, A. P. {\it et al.}, NMR Evidence for Inhomogeneous Nematic Fluctuations in
BaFe$_2$(As$_{1-x}$P$_x$)$_2$. Phys. Rev. Lett. {\bf 116},107202 (2016).

\bibitem{fradkin} Fradkin, E., Kivelson, S.A., Lawler, M.J., Eisenstein, J.P. $\&$ Mackenzie, A.P. Nematic Fermi Fluids in Condensed Matter Physics. Annual Review of Condensed Matter Physics {\bf 1}, 153-178 (2010).

\bibitem{CFang08} Fang, C., Yao, H., Tsai, W.-F., Hu, J. $\&$ Kivelson, S.A. Theory of electron nematic order in LaFeAsO. Phys. Rev. B {\bf 77}, 224509 (2008).

\bibitem{Fernandes11} Fernandes, R.M., Abrahams, E. $\&$ Schmalian, J. Anisotropic in-plane resistivity in the nematic phase of the iron pnictides. Phys. Rev. Lett. {\bf 107}, 217002 (2011).

\bibitem{Fernandes14} Fernandes, R.M., Chubukov, A.V. $\&$ Schmalian, J. What drives nematic order in iron-based superconductors? Nature Physics {\bf 10}, 97-104 (2014).

\bibitem{chu10} Chu, J. H. {\it et al.} In-Plane Resistivity Anisotropy in an Underdoped Iron Arsenide Superconductor. Science {\bf 329}, 824-826 (2010).

\bibitem{matanatar} Tanatar, M. A. {\it et al.} Uniaxial-strain mechanical detwinning of CaFe$_2$As$_2$ and BaFe$_2$As$_2$ crystals: Optical and transport study. Phys. Rev. B {\bf 81}, 184508 (2010).

\bibitem{fisher} Fisher, I.R., Degiorgi, L. $\&$ Shen, Z.X. In-plane electronic anisotropy of underdoped '122' Fe-arsenide superconductors revealed by measurements of detwinned single crystals. Reports on Progress in Physics {\bf 74}, 124506 (2011).

\bibitem{Kasahara} Kasahara, S. et al. Electronic nematicity above the structural and superconducting transition in BaFe$_2$(As$_{1-x}$P$_x$)$_2$. Nature {\bf 486}, 382 (2012).

\bibitem{xyscience} Lu, X. Y. {\it et al.} Nematic spin correlations in the tetragonal state of uniaxial-strained BaFe$_{2-x}$Ni$_x$As$_2$. Science {\bf 345}, 657 (2014).

\bibitem{Rosenthal} Rosenthal, E.P. {\it et al.} Visualization of electron nematicity and unidirectional antiferroic fluctuations at high temperatures in NaFeAs. Nat. Phys. {\bf 10}, 225-232 (2014).

\bibitem{Kuo2016} Kuo, H. H., Chu, J. H., Palmstrom, J. C., Kivelson, S. A., Fisher, I. R., Ubiquitous signatures of nematic quantum criticality in optimally doped Fe-based superconductors. Science {\bf 352}, 958 (2016).

\bibitem{Analytis} Analytis, J. G. {\it et al.} Transport near a quantum critical point in BaFe$_2$(As$_{1-x}$P$_x$)$_2$. Nat. Phys. {\bf 10}, 194-197 (2014).

\bibitem{Metlitski} Metlitski, M. A., Mross, D. F., Sachdev, S., \& Senthil, T., Cooper pairing in non-Fermi liquids. Phys. Rev.
B {\bf 91}, 115111 (2015).

\bibitem{Lederer} Lederer, S., Schattner, Y., Berg, E., Kivelson, S. A., Enhancement
of superconductivity near a nematic quantum critical point. Phys. Rev. Lett. {\bf 114}, 097001 (2015).

\bibitem{zhao08} Zhao, J. {\it et al.} Structural and magnetic phase diagram of CeFeAsO$_{1- x}$F$_x$ and its relation to high-temperature superconductivity. Nat. Mater. {\bf 7}, 953 (2008).

\bibitem{lee08} Lee, C.-H. {\it et al.} Effect of Structural Parameters on Superconductivity in Fluorine-Free LnFeAsO$_{1-y}$ (Ln = La, Nd). Journal of the Physical Society of Japan {\bf 77}, 083704 (2008).

\bibitem{kote08} Kotegawa, H., Kawazoe, T., Sugawara, H., Murata, K. $\&$ Tou, H. Effect of Uniaxial Stress for Pressure-Induced Superconductor SrFe$_2$As$_2$. Journal of the Physical Society of Japan {\bf 78}, 083702 (2009).

\bibitem{kuro09} Kuroki, K., Usui, H., Onari, S., Arita, R. $\&$ Aoki, H. Pnictogen height as a possible switch between high-Tcnodeless and low-Tcnodal pairings in the iron-based superconductors. Phys. Rev. B {\bf 79}, 224511 (2009).

\bibitem{xianhui14} Chen, X., Dai, P., Feng, D., Xiang, T. $\&$ Zhang, F.-C. Iron-based high transition temperature superconductors. National Science Review {\bf 1}, 371-395 (2014).

\bibitem{kasa10} Kasahara, S. {\it et al.}, Evolution from non-Fermi- to Fermi-liquid transport via isovalent doping in BaFe$_2$(As$_{1-x}$P$_x$)$_2$ superconductors. Phys. Rev. B {\bf 81}, 184519 (2010).

\bibitem{bohmer12} B$\rm \ddot{o}$hmer, A. E. {\it et al.}, Thermodynamic phase diagram, phase competition, and uniaxial pressure effects in BaFe$_2$(As$_{1-x}$P$_x$)$_2$ studied by thermal expansion. Phys. Rev. B {\bf 86}, 094521 (2012).

\bibitem{fisher2012} Kuo, H. H. {\it et al.}, Magnetoelastically coupled structural, magnetic, and superconducting order parameters in BaFe$_2$(As$_{1-x}$P$_x$)$_2$. Phys. Rev. B {\bf 86}, 0134507 (2012).

\bibitem{Hirschfeld2011} Hirschfeld, P. J., Korshunov, M. M. \& Mazin, I. I. Gap symmetry and structure of Fe-based superconductors. Reports on Progress in Physics {\bf 74}, 124508 (2011).

\bibitem{rotter10} Rotter, M., Hieke, Ch., and Johrendt, D., Different response of the crystal structure to isoelectronic doping in BaFe$_2$(As$_{1-x}$P$_x$)$_2$ and (Ba$_{1-x}$Sr$_x$)Fe$_2$As$_2$. Phys. Rev. B {\bf 82}, 014513 (2010).

\bibitem{DMFT-RMP2006} Kotliar, G. {\it et al.}, Electronic structure calculations with dynamical mean-field theory. Reviews of Modern Physics {\bf 78}, 865-951 (2006).

\bibitem{analy10} Analytis, J.G., Chu, J.H., McDonald, R.D., Riggs, S.C. $\&$ Fisher, I.R. Enhanced Fermi-surface nesting in superconducting BaFe$_2$(As$_{1-x}$P$_x$)$_2$ revealed by the de Haas-van Alphen effect. Phys. Rev. Lett. {\bf 105}, 207004 (2010).

\bibitem{beek10} van der Beek, C.J. {\it et al.}, Quasiparticle scattering induced by charge doping of iron-pnictide superconductors probed by collective vortex pinning. Phys. Rev. Lett. {\bf 105}, 267002 (2010).

\bibitem{klin10} E. Klintberg, L. {\it et al.}, Chemical Pressure and Physical Pressure in BaFe$_2$(As$_{1-x}$P$_x$)$_2$. Journal of the Physical Society of Japan {\bf 79}, 123706 (2010).

\bibitem{dhital} Dhital, C. {\it et al.}, Effect of uniaxial strain on the structural and
magnetic phase transitions in BaFe$_2$As$_2$. Phys. Rev. Lett. {\bf 108}, 087001 (2012).

\bibitem{Dhital14} Dhital, C. {\it et al.}, Evolution of antiferromagnetic susceptibility under uniaxial pressure in Ba(Fe$_{1-x}$Co$_x$)$_2$As$_2$. Phys. Rev. B {\bf 89}, 214404 (2014).

\bibitem{YSong13} Song, Y. {\it et al.}, Uniaxial pressure effect on structural and magnetic phase transitions in NaFeAs and its comparison with as-grown and annealed BaFe$_2$As$_2$. Phys. Rev. B {\bf 87}, 184511 (2013).

\bibitem{man} Man, H. R. {\it et al.}, Electronic nematic correlations in the stress-free tetragonal state of BaFe$_{2?x}$Ni$_x$As$_2$.
Phys. Rev. B {\bf 92}, 134521 (2015).

\bibitem{XYLu15} Lu, X. Y. {\it et al.}, Impact of uniaxial pressure on structural and magnetic phase transitions in electron-doped iron pnictides. Phys. Rev. B {\bf 93}, 134519 (2016).

\bibitem{SI} See supplementary information for details of transport and neutron scattering experiments.

\bibitem{Tam17} Tam, D.W. {\it et al.}, Uniaxial pressure effect on the magnetic ordered moment and transition temperatures in BaFe$_{2-x}$T$_x$As$_2$ (T=Co,Ni). Phys. Rev. B {\bf 95}, 060505(R) (2017).

\bibitem{robert15} Georgii, R. \& Seemann, K. MIRA: Dual wavelength band instrument.
Journal of large-scale research facilities {\bf 1}, A3 (2015).

\bibitem{ZPY08} Yin, Z. P. {\it et al.}, Electron-hole symmetry and magnetic coupling in antiferromagnetic LaFeAsO.
Phys. Rev. Lett. {\bf 101}, 047001 (2008).

\bibitem{ZPY11_2} Yin, Z.P., Haule, K. \& Kotliar, G. Kinetic frustration and the nature of the magnetic and paramagnetic states in iron pnictides and iron chalcogenides. Nat. Mater. {\bf 10}, 932-5 (2011).

\bibitem{Haule-QMC} Haule, K. Quantum Monte Carlo impurity solver for cluster dynamical mean-field theory and electronic structure calculations with adjustable cluster base. Phys. Rev. B {\bf 75}, 155113 (2007).

\bibitem{Werner} Werner, P., Comanac, A., De' Medici, L., Troyer, M. \& Millis, A. J. Continuous-time solver for quantum impurity models.
Phys. Rev. Lett. {\bf 97}, 076405 (2006).

\bibitem{ZPY11} Yin, Z.P., Haule, K. \& Kotliar, G. Magnetism and charge dynamics in iron pnictides. Nat. Phys. {\bf 7}, 294-297 (2011).

\bibitem{ZPY13} Yin, Z. P., Kutepov, A. \& Kotliar, G. Correlation-Enhanced Electron-Phonon Coupling: Applications of GWand Screened Hybrid Functional to Bismuthates, Chloronitrides, and Other High-$T_c$ Superconductors. Phys. Rev. X {\bf 3}, 021011 (2013).

\bibitem{Mandal14} Mandal, S., Cohen, R. E., \& Haule, K. Strong pressure-dependent electron-phonon coupling in FeSe. Phys. Rev. B {\bf 89}, 220502(R) (2014).

\bibitem{Gerber17} Gerber, S. {\it et al.} Femtosecond electron-phonon lock-in by photoemission and x-ray free-electron laser. Science {\bf 357}, 71 (2017).

\bibitem{fujii2018} Chiaki Fujii {\it et al.}, Diverse fluctuations and anisotropic Gruneisen parameter behavior in iron-based superconductor Ba(Fe$_{1-x}$Co$_x$)$_2$As$_2$ and their correlation with superconductivity. arxiv {\bf 1801},02791.

\bibitem{bohmer17} B$\rm \ddot{o}$hmer, A. E. {\it et al.}, Effect of Biaxial Strain on the Phase Transitions of Ca(Fe$_{1-x}$Co$_x$)$_2$As$_2$. Phys. Rev. Lett. {\bf 118}, 107002 (2017).

\bibitem{Ca122-Boehmer2} Fente, A. {\it et al.} Direct visualization of phase separation between superconducting and nematic domains in Co-doped CaFe$_2$As$_2$ close to a first-order phase transition. {\it Phys. Rev. B} {\bf 97,} 014505 (2018).



\end{thebibliography}
\end{document}